\documentstyle[aps,prl,epsfig]{revtex}

\begin{document}

\draft

\title{Radiation Pressure Induced Einstein-Podolsky-Rosen Paradox}

\author{
Vittorio Giovannetti${}^1$,
Stefano Mancini${}^2$
and Paolo Tombesi${}^1$
}

\address{
${}^1$INFM, Dipartimento di Matematica e Fisica, Universit\`a di
Camerino, I-62032 Camerino, Italy\\
${}^2$INFM, Dipartimento di Fisica,
Universit\`a di Milano,
Via Celoria 16, I-20133 Milano, Italy
}

\date{\today}

\maketitle

\widetext

\begin{abstract}
We demonstrate the appearance
of Einstein-Podolsky-Rosen (EPR) paradox when 
a radiation field impinges on a movable mirror.
Then, the possibility of a local realism test 
within a pendular Fabry-Perot cavity is 
shown to be feasible.  
\end{abstract}


Quantum Optics is widely used to 
perform fundamental tests of Quantum Mechanics.
Essentially, the reason relies on the fact that the only relevant noise 
at optical frequency is the vacuum noise, and on the possibility
of generating non-classical states through nonlinear media.
Recently, however, much of theoretical attention has also been devoted to
opto-mechanical systems \cite{OPTO}.
On the other hand, due to technological developments, 
those systems are now becoming experimentally accessible \cite{EXP}.

Along this line, we would deal with the interesting question
of whether radiation pressure can induce entanglement.
Entanglement is, indeed, the source of many mysteries of Quantum 
Mechanics and it is the fundamental concept involved in what is called 
the spooky action-at-a-distance \cite{EPR}. In this respect our 
analysis will show that entanglement can be obtained via a 
macroscopic object. It is not quite entanglement between macroscopic
objects but it will be shown that it is mediated by a macroscopic body.
To this end we consider the radiation
field impinging on a movable mirror and 
we will study the properties of the reflected field.
The quantum noise reduction for such a model has been predicted \cite{STEFANO1},
however, we demonstrate here that the state of the radiation field 
interacting with a macroscopic object can become non-classical at
a level deeper than simple squeezing. Namely, we demonstrate 
the appearance of EPR aspects \cite{EPR} on continuous variables in that system.

Our analysis starts from the EPR paradox theoretically formulated
in Refs. \cite{REID1,REID2}, where the roles of canonical position 
and momentum variables, of the original EPR 
paper \cite{EPR}, are played by the amplitude and phase quadratures 
of the output
signals belonging to two distinct cavity modes.
In this scheme the quadratures $\{\hat{X}_{1}, 
\hat{Y}_{1}\}$ of the 
first mode are inferred in turn from measurements of the
quadratures $\{\hat{X}_{2},\hat{Y}_{2}\}$ of 
the second mode, which is let 
spatially separate from the first.
The errors of these inferences are then quantified by the variances
\begin{equation}
\Delta_{\rm inf}^{2} \hat{X}(g_{X}) \equiv \left\langle \left( \hat{X}_{1} -
g_{X} 
\hat{X}_{2} \right)^{2} \right\rangle  \hspace{.5 in} 
\Delta_{\rm inf}^{2} \hat{Y}(g_{Y}) \equiv \left\langle \left( \hat{Y}_{1} -
g_{Y} 
\hat{Y}_{2} \right)^{2} \right\rangle
\label{ERROR}
\end{equation}
where $g_{X}$ and $g_{Y}$ are two dimensionless scaling factors, which 
we will adjust to allow for the greatest accuracy in inferred 
determinations of $\{\hat{X}_{1}, \hat{Y}_{1}\}$.
As pointed out in Ref. \cite{REID2}, an experimental
demonstration of the EPR paradox occurs when
the product of the errors of the inferences will be less than the
limit imposed by the Heisenberg principle for 
the observables $\{\hat{X}_{1},\hat{Y}_{1}\}$.
Of course, in order to obtain such a result the modes have to be 
strongly correlated. In the original proposal of Refs.
\cite{REID1,REID2} and in the experimental realization of Ref. 
\cite{KIMBLE},
the necessary correlation was obtained by a 
nondegenerate optical parametric oscillator, which 
makes interact the two modes inside the cavity.

Here, instead,
they indirectly interact, by means of the radiation pressure 
by which each of them acts on one of the cavity mirrors,
which is let free to oscillate and assumed perfectly reflecting.
The other mirror is instead assumed fixed and with non zero transmittivity.
When the cavity is empty the moving mirror 
undergoes harmonic oscillations damped by the 
coupling to an external bath in equilibrium at temperature $T$.
As pointed out in Ref.  \cite{BRAG}, under the assumption 
that the measurement time is of the order of the mechanical relaxation 
time of the mirror, it is possible to consider such a macroscopic 
oscillator as  a quantum oscillator. The cavity resonances are 
calculated in the absence of the impinging field, hence if $L$ is the 
equilibrium cavity length the resonant frequency of the cavity will be
$\omega_{c}=nc/2L$, where $n$ is an arbitrary integer number 
and $c$ the speed of light. Furthermore, we assume
that at the frequency of the impinging field $\omega_{0}$, 
the fixed mirror does 
not introduce any excess noise beyond the input field noise. We also 
assume that retardation effects, due to
the oscillating mirror in the intracavity field, are negligible.
We shall use a field intensity such that the correction to the 
radiation pressure force, due to the Doppler frequency shift of 
the photons \cite{UNRU} on the moving mirror, is completely negligible.
This means considering the damping coefficient of the oscillating 
mirror to be only due to the coupling with the thermal bath.
Thus, the Hamiltonian for this system can be written as
\begin{equation}
{\hat H}=\hbar \omega_{c} \left( {\hat A}_{1}^{\dagger}{\hat A}_{1} 
+ {\hat A}_{2}^{\dagger}{\hat A}_{2} + 1
\right)  +
 \frac{\hat{P}^{2}}{2 m} + \frac{1}{2}m \, 
 \omega_{m} \hat{Q}^{2} + {\hat H}_{int}
\label{H}
\end{equation}
where ${\hat A}_{1}$ and ${\hat A}_{2}$ are the annihilation operators 
of the two cavity modes, which are supposed to have the same frequency 
and mutually orthogonal polarizations; 
$\hat{P}$ and $\hat{Q}$ are the momentum and 
the displacement operators, respectively, from the equilibrium position 
of the oscillating mirror with mass $m$ and oscillation frequency 
$\omega_{m}$. The mechanical frequency $\omega_{m}$ will be 
many orders of magnitude smaller than $\omega_{c}$ to ensure that the 
number of photons generated by the Casimir effect is completely 
negligible \cite{CASIMIR}. ${\hat H}_{int}$ accounts for the interaction 
between the cavity modes and the oscillating mirror \cite{LAW}. 
Since we have assumed no retardation effects, ${\hat H}_{int}$ simply 
represents the effect of the radiation pressure force which causes the 
instantaneous displacement $\hat{Q}$ of the mirror \cite{STEFANO1,PACE}:
\begin{equation}\label{Hint}
{\hat H}_{int}=-\hbar \omega_{c} \, \left( {\hat A}_{1}^{\dagger}{\hat A}_{1} +
{\hat A}_{2}^{\dagger}{\hat A}_{2}\right)  \;  \hat{Q}  /L \; .
\label{hint} 
\end{equation}
Now, we assume that the intracavity radiation fields 
are damped through the output fixed
mirror at the same rate $\gamma_c$, 
while $\gamma_m$ is the mechanical damping rate for the 
mirror's Brownian motion. 
The interaction (\ref{Hint}) gives rise to nonlinear 
Langevin equations whose linearization around the steady state leads to
\begin{equation}
\left\{
\begin{array}{l}
\frac{d}{dt} \, \hat{q}(t) = \frac{1}{m}\hat{p}(t) \,,  \\ \\
\frac{d}{dt} \, \hat{p}(t) = -m \, \omega_{m}^{2} \hat{q}(t) -  
  2 \, \gamma_m \hat{p}(t) 
  +\hbar \omega_{c} \frac{|\alpha_{1}|}{L} \left( {\hat a}_{1}(t) + 
  {\hat a}_{1}^{\dagger}(t) \right)  +
  \hbar \omega_{c} \frac{|\alpha_{2}|}{L} \left( {\hat a}_{2}(t) + 
  {\hat a}_{2}^{\dagger}(t)\right)
  -  {\hat\xi}(t) \,,
  \\ \\
\frac{d}{dt} \, {\hat a}_{j}(t) =  \left(i \Delta -\frac{1}{2}\right) \gamma_{c}
{\hat a}_{j}(t) + i 
\omega_{c} \frac{|\alpha_{j}|}{L} \hat{q}(t) + {\hat a}_{j}^{\rm in}(t) \,,
\hspace{.5 in} 
 \hspace{.1 in} j=1,2  
\end{array} \right.
\label{heisenberg}
\end{equation}
where now all the operators represent small fluctuations
around steady state values, i.e.
$\hat{Q}(t)=x+\hat{q}(t)$,
$\hat{P}(t)=y+\hat{p}(t)$,
${\hat A}_{j}(t)=\big( \alpha_j+{\hat a}(t) \big) \exp[- i(\omega_{0}t -
\arg{\alpha_{j}})]$. The steady state is determined by 
\begin{equation}
\begin{array}{llll}
x \equiv \langle {\hat Q} \rangle_{ss}
= \frac{\hbar \omega_{c}}{m \omega_{m}^{2} L} \left( | \alpha_{1}|^{2}
+| \alpha_{2}|^{2} \right) \,, & & & 
y \equiv \langle {\hat P} \rangle_{ss} = 0 \,, \\
\\
\alpha_{j} \equiv  \langle {\hat A}_j \rangle_{ss} e^{i \omega_{0}t}
= \frac{\sqrt{\gamma_{c}} \alpha_{j}^{in}}{(1/2 -i
\Delta)\gamma_{c}} \,, & & &
\Delta = (\omega_{0} - \omega_{c} + \omega_{c}x/L)/\gamma_c \,, \\
\end{array}
\label{costanti}
\end{equation}
where $\alpha_{j}^{in}$ is the classical field characterizing the input 
laser power
$P_{in}= \hbar \omega_{0} \left(|\alpha_{1}^{in}|^{2}
+|\alpha_{2}^{in}|^{2} \right)$.
In the following we assume that the laser pumps both the 
cavity modes in the same way, i.e. $\alpha_{1}^{in} = \alpha_{2}^{in}$: 
this allows for identical coupling constants between the two modes 
and the mirror, $\alpha_{1}=\alpha_{2}=\alpha$.
Moreover, $\Delta$ is the (dimensionless) 
radiation phase shift due to the detuning and
to the stationary displacement of the mirror. 
${\hat a}^{\rm in}_{j}(t)$ is the vacuum noise operator 
at the input of the $j${\em-th} cavity  mode.
Instead,
${\hat\xi}(t)$ is the noise operator for the
quantum Brownian motion 
\cite{DIOSI} of the mirror.
The non vanishing noise correlations are
\begin{eqnarray}\label{noise}
\langle {\hat a}^{\rm in}_j(t) [{\hat a}^{\rm in}_k]^{\dag}(t') \rangle &=& 
\gamma_c \delta(t-t')\,\delta_{j,k}\,,\quad j,k=1,2\,,
\\
\langle {\hat\xi}(t) {\hat\xi}(t') \rangle &=&
\frac{m \gamma_{m}}{\pi} \int \, d\omega \, \left\{ 
\hbar\omega\left[\coth\left(
\frac{\hbar\omega}{2k_BT}\right)-1 \right] \right\} 
e^{i\omega(t-t')}\,,
\end{eqnarray}
where $k_B$ is the Boltzmann constant and $T$ the equilibrium temperature.

The boundary relation for the radiation fields is \cite{GARDINER}
\begin{equation}\label{inout}
{\hat a}^{\rm out}_j(t) = \gamma_c \, {\hat a}_j(t)
-{\hat a}^{\rm in}_j(t)\,.
\end{equation}

From Eq. (\ref{heisenberg}) we can see that the cavity modes ${\hat a}_{1}(t)$ 
and ${\hat a}_{2}(t)$, mutually interact by means of the
action of the mirror displacement operator. In particular this
coupling is linear and it seems to be able to create correlation
between the modes quadratures, 
which in turn can be revealed in the output field. 
At the beginning of this letter 
we have adopted an heuristic approach in order to introduce
the problem. We considered the quadrature operators of the signal 
modes as effective measurable quantities,
which  obey the canonical commutation 
rules that permit us to write down an Heisenberg's relation
by which testing the EPR paradox. 
However, in the case of a signal outgoing from a cavity
one has to be careful, since 
the commutation rules at two distinct time is 
a Dirac delta function of the difference of the time values 
considered \cite{GARDINER}: this means that it is impossible to write down an
effective uncertainty relation for their quadratures.
The solution for this problem is to consider such operators 
in the frequency domain 
instead of the time domain \cite{REID1,REID2}. 

Thus, given the quadrature 
\begin{equation}\label{quadef}
\hat{X}^{\rm out}_{j}(\phi, t)=
{\hat a}^{\rm out}_j(t) e^{-i\phi}
+[{\hat a}^{\rm out}_j(t)]^{\dag}(t) e^{i\phi}\,,
\end{equation}
we define
\begin{equation}
\hat{{\widetilde X}}^{\rm out}_{j}(\phi,\omega) 
= \frac{1}{\sqrt{\tau}} \int_{-\tau/2}^{\tau/2} 
\, dt \,
e^{i \omega t} \, \hat{X}^{\rm out}_{j}(\phi, t) \; ,
\label{operatore}
\end{equation}
where $\tau$ is the measurement time of the output 
signal.
$\hat{\widetilde X}^{\rm out}_{j}(\phi,\omega)$ is a sort of Fourier transform
of $\hat{X}^{\rm out}_{j}(\phi, t)$ on a limited time interval, which
has some interesting properties. For example, in the limit of $\tau$ 
greater than the coherence time of the signal,
the expectation values of the product of 
$\hat{\widetilde X}^{\rm out}_{j}(\phi,\omega)$  
with $\hat{\widetilde X}^{\rm out}_{k}(\varphi,\omega)$  is the power density
spectrum of the original quadrature fluctuations product, namely
\begin{equation}
S_{j,k}(\omega,\phi,\varphi) = \left\{ \int dt' e^{i \omega t'} 
\langle \; \hat{X}^{\rm out}_{j}(\phi,t) \; 
 \hat{X}^{\rm out}_{k}(\varphi,t'-t)\rangle \right\}_{t} \; ,
\label{power}
\end{equation}
where $\{ \cdots \}_{t}$ means averaging $t$ on the time interval 
$\tau$. $S_{j,k}(\omega,\phi,\varphi)$ is simply
related to the spectral density photocurrent fluctuations, which is 
possible to measure by means of separated balanced homodyne detection
of the two signal (see for example Ref. \cite{KIMBLE}).
In the rest we focus on the case with $\omega=0$. Notice that this is the
only case in which the operators (\ref{operatore}) are Hermitian.  
Furthermore, the commutation relation between 
$\hat{\widetilde X}^{\rm out}_{j}(\phi,0)$ and
$\hat{\widetilde X}^{\rm out}_{k}(\phi+\pi/2,0)$ is exactly equal to $2 i 
\gamma_{c} \delta_{j,k}$ as it is possible to verify from the 
properties of the output signals. This means that
$\{\hat{\widetilde X}^{\rm out}_{1}(0,0) , 
\hat{\widetilde X}^{\rm out}_{1}(\pi/2,0) \}$ 
and $\{\hat{\widetilde X}^{\rm out}_{2}(0,0) , 
\hat{\widetilde X}^{\rm out}_{2}(\pi/2,0) \}$ 
are two mutually independent couples of conjugate observables.
The only constraint in order to use them in an EPR paradox 
realization, is to ensure that the measurements on the 
first mode are spatially separated from those of the second mode. In 
particular this means $c \tau /d < 1$, with  
$d$ the spatial distance between the balanced homodyne
receivers of the modes. 

Now, by Fourier transforming Eqs. (\ref{heisenberg}) it is possible to 
get the output field quadratures by means of Eq.(\ref{inout})
and then one can evaluate the inferences errors (\ref{ERROR}).
Choosing of inferring $\hat{\widetilde X}^{\rm out}_{1}(0,0)$ and 
$\hat{\widetilde X}^{\rm out}_{1}(\pi/2,0)$ with 
$\hat{\widetilde X}^{\rm out}_{2}(0,0)$ and 
$\hat{\widetilde X}^{\rm out}_{2}(\pi/2,0)$ respectively,
we obtain
\begin{equation}
\Delta^2_{\rm inf} \hat{\widetilde X}^{\rm out}_1(\phi,0) 
=\sum_{j,k=1}^2 (-)^{j+k}\, g_{\phi}^{j+k-2} \,
S_{j,k}(0,\phi,\phi)\,,
\end{equation}
and minimizing the inference errors by the scaling factors 
$g_{\phi}$ introduced in Eq. (\ref{ERROR}), the above equation reduces to
\begin{equation}
\Delta_{\rm inf}^{2} \hat{\widetilde X}^{\rm out}_1(\phi,0) 
= \gamma_{c} \left( 
1 + \frac{\epsilon(\phi)}{1 + \epsilon(\phi)} \right) \,,
\label{delta}
\end{equation}
with
\begin{equation}
\epsilon(0)= \left( \frac{{\cal T}-1}{2} \right)
\frac{\cal P}{\left( {\Delta}^{2} + {\cal P} + 1/4 \right)^{2}} \; ,
\hspace{.5 in} 
\epsilon(\pi/2)=\left(\frac{{\Delta}^{2}+{\cal P}
+{\cal T}/4}{2 {\Delta}^{2}}\right)
\frac{\cal P}{\left( {\Delta}^{2} + {\cal P} + 1/4 \right)^{2}} \; ,
\label{epsilon}
\end{equation}
where we have introduced the dimensionless power and temperature 
${\cal P}= 8 \omega_{0} {\Delta} P_{in} / m 
L^{2} \omega_{m}^{2} \gamma_{c}^{2}(1+4\Delta^{2})$,
${\cal T}=8 k_B T \gamma_m \Delta / \hbar \omega_{m}^{2}$.
These become now the fundamental variables of our system. 

The EPR paradox will be verified if the product of the inference
errors $\Delta_{\rm inf}^{2} \hat{\widetilde X}^{\rm out}(0,0)$ and 
$\Delta_{\rm inf}^{2} \hat{\widetilde X}^{\rm out}(\pi/2,0)$ is less than the 
minimum uncertainty limit imposed by the commutation relation between
$\hat{\widetilde X}^{\rm out}_{1}(0,0)$ and 
$ \hat{\widetilde X}^{\rm out}_{1}(\pi/2,0)$;
that means, with the aid of Eq. (\ref{delta}),
\begin{equation}
\left( 
1 + \frac{\epsilon(0)}{1 + \epsilon(0)} \right) \left( 
1 + \frac{\epsilon(\pi/2)}{1 + \epsilon(\pi/2)} \right) < 1 \; .
\label{epr1}
\end{equation}
From Eq. (\ref{epsilon}) results that $\epsilon(\pi/2)$ is 
always positive; so, the term in the second brackets on the l.h.s. of Eq. 
(\ref{epr1})
is always greater than $1$. Then, we can verify the EPR paradox  
if the term in the first brackets on the l.h.s. of Eq. (\ref{epr1}) 
becomes smaller than 1.
That depends on $\epsilon(0)$ which, for ${\cal T}<1$, can be a
negative number. 

In Fig.1 we show the l.h.s. of Eq.(\ref{epr1})
as function of ${\cal P}$ and ${\cal T}$. We see 
how the thermal noise related to the Brownian motion 
of the mirror plays a detrimental role,
in fact as soon as ${\cal T}$ increases, the region of possible values of 
${\cal P}$ tends to become narrower and to disappear.
Notice that writing back $P_{in}$ in terms of ${\cal P}$, the limit 
$\Delta\to 0$ cannot be performed;
this is consistent with Eqs.(\ref{heisenberg}) 
where in such a limit only the amplitude 
quadratures become correlated, hence no paradox can arise.

For grant, we take from the graph ${\cal P}=0.17$ and ${\cal T}=0.1$
which can be obtained with
the following values of parameters:
$m = 3 \times 10^{-5}$  Kg, $L = 10^{-3}$ m,
$\omega_{m} = 2$ MHz, $\gamma_m = 1$ Hz,
$\omega_{c} \simeq \omega_{0}= 2 \times 10^{15} $ Hz,
$\gamma_{c}= 2$ MHz,
$T \simeq 4 \; ^{\circ}{\mbox K}$ and 
$P_{in} \simeq 30  \mbox{mW}$. 
These correspond to the experimental set up
of Ref. \cite{COHADON}, and give the value $0.7$ 
for the l.h.s. of Eq.(\ref{epr1}).
Of course, better results can be achieved by means of an optimization
of all parameters or by
working with microfabricated cantilever \cite{CLELAND}.

In conclusion, we have proposed a {\it ponderomotive} scheme to get 
entanglement of radiation fields. The resulting correlations
give rise to EPR paradox and seem quite robust notwithstanding
they derive from the action of a macroscopic object.
Furthermore, the numerical results show that the proposed test can be 
realized with the present technology.

\begin{figure}[t]
\centerline{\epsfig{figure=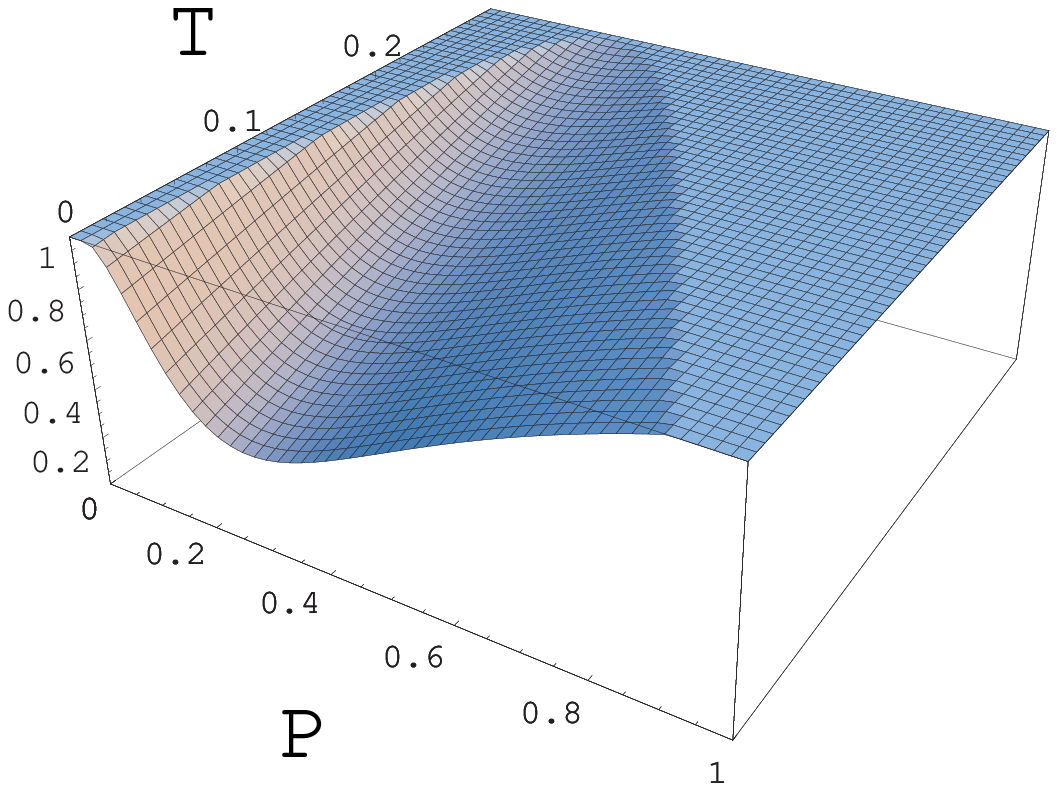,width=3.5in}}
\caption{\widetext 
The l.h.s. of Eq.(\ref{epr1}) is plotted vs. ${\cal P}$ and
${\cal T}$ for $\Delta=0.18$.
}
\label{fig1}
\end{figure}


\begin{thebibliography}{99}

\bibitem{OPTO}
See, e.g., S. Mancini, {\it et al.}, Phys. Rev. Lett. {\bf 80}, 688 (1998);
S. Bose, {\it et al.}, Phys. Rev. A {\bf 56}, 4175 (1997);
S. Bose, {\it et al.}, Phys. Rev. A {\bf 59}, 3204 (1999);
K. Jacobs {\it et al.}, Phys. Rev. A {\bf 60}, 538 (1999).

\bibitem{EXP}
I. Tittonen, {\it et al.}, Phys. Rev. A {\bf 59}, 1038 (1999);
Y. Hadjar, {\it et al.}, Europhys. Lett. {\bf 46}, 545 (1999);
P. F. Cohadon, {\it et al.}, Phys. Rev. Lett. {\bf 83}, 3174 (1999).

\bibitem{EPR}
A. Einstein, {\it et al.}, Phys. Rev. {\bf 47}, 777 (1935).

\bibitem{STEFANO1}
S. Mancini and P. Tombesi, Phys. Rev. A {\bf 49}, 4055 (1994); 
C. Fabre, {\it et al.}, Phys. Rev. A {\bf 49}, 1337 (1994).

\bibitem{REID1}
M.D. Reid and P.D. Drummond, Phys. Rev. A {\bf 40}, 
4493 (1989); Phys. Rev. A {\bf 41}, 3930 (1990).

\bibitem{REID2}
M. D. Reid, Phys. Rev. A, {\bf 40} 913, (1992).

\bibitem{KIMBLE}
Z.Y. Ou, {\it et al.}, Phys. Rev. Lett. {\bf 68}, 3663 (1992). 

\bibitem{BRAG}
V.B. Braginsky and F.Ya. Khalili, {\em Quantum 
Measuraments}, Ed. by K.S. Thorne (Cambridge Univ. Press, 
Cambridge, 1992); V.B. Braginsky and V.P. Mitrofanov, {\em Systems With 
Small Dissipation} (Univ. of Chicago Press, Chicago, 1985).

\bibitem{UNRU}
W. Unruh, in {\em Quantum Optics, Experimental 
Gravitation and Measuraments Theory}, Ed. by P. Meystre and M. O. 
Scully (Plenum, New York, 1983).

\bibitem{CASIMIR}
H. Casimir, Proc. Kon. Ned. Akad. Wet. {\bf 51}, 793 (1948).

\bibitem{LAW}
C.K. Law, Phys. Rev. A {\bf 51}, 2537 (1995).

\bibitem{PACE}
A.F. Pace {\em et al.}, Phys. Rev. A {\bf 47}, 3173 (1993).

\bibitem{DIOSI}
L. Di\'osi, Physica A {\bf 199}, 517 (1993).

\bibitem{GARDINER}
C. W. Gardiner, {\em Quantum noise} (Springer-Verlag, Berlin, 1991).

\bibitem{COHADON}
P.F. Cohadon, {\it et al.}, Phys. Rev. Lett. {\bf 83}, 3174 (1999).

\bibitem{CLELAND}
A.N. Cleland and M.L. Roukes, Appl. Phys. Lett. {\bf 69}, 2653 (1996).

\end{thebibliography}
\end{document}